\documentclass{article}
\usepackage{amsmath,latexsym}
\usepackage{amssymb}
\usepackage{spconf,amsmath,epsfig}
\usepackage{url}
\usepackage{epic,eepic}

\newcommand{\refeq}[1]{(\ref{#1})}
\newcommand{\E}{\ensuremath{{\mathbb E}}}
\newcommand{\ms}{\;\;}
\newcommand{\mycaption}[1]{\vspace*{-0.75em}\caption{#1}\vspace*{-0.5em}}
\newcommand{\mysection}[1]{\vspace*{-0.25em}\section{#1}\vspace*{-0.15em}}

\ninept
\title{A Unified Approach to Joint and Iterative Adaptive Interference Cancellation and Parameter Estimation
for CDMA Systems in Multipath Channels \vspace*{-0.0em}}
\name{ Rodrigo C.\ de Lamare, Raimundo Sampaio-Neto, and Are
  Hj{\o}rungnes \thanks{This work was supported by the Research
    Council of Norway VERDIKT project 176773/S10 called OptiMO.}\vspace*{-0.1em}}
\address{\normalsize Communications Research Group, Department of Electronics, University of York, United Kingdom\\
  \normalsize CETUC, Pontifical Catholic University of Rio de Janeiro
  (PUC-RIO),
  Brazil \\
  \normalsize UniK - University Graduate Center, University of Oslo, Norway\\
  {\normalsize Emails: \url{rcdl500@ohm.york.ac.uk,
      raimundo@cetuc.puc-rio.br, arehj@unik.no}}
  \vspace*{-0.0em}}

\begin{document}
%

\maketitle
\vspace*{-1.5em}
\begin{abstract}{ 
    This paper proposes a unified approach to joint adaptive
    parameter estimation and interference cancellation~(IC) for direct sequence
    code-division-multiple-access~(DS-CDMA) systems in multipath channels. A unified framework is presented in which the
    IC problem is formulated as an optimization problem with extra degrees of freedom of an IC
    parameter vector for each stage and user. We propose a joint optimization
    method for estimating the IC parameter vector, the linear receiver
    filter front-end, and the channel along with minimum
    mean squared error~(MMSE) expressions for the estimators. Based on
    the proposed joint optimization approach, we derive low-complexity
    stochastic gradient~(SG) algorithms for estimating the desired
    parameters. Simulation results for the uplink of a synchronous
    DS-CDMA system show that the proposed methods significantly
    outperform the best known IC receivers.}
\end{abstract}
\vspace*{-0.1em}
\mysection{Introduction} \label{sec:intro}

High data rate applications for future wireless systems require an
ever-increasing sophistication and performance of receivers. In
multi\-user systems such as DS-CDMA, novel signal processing techniques are
of crucial importance to enhance the capacity and the performance. The
field of interference mitigation techniques has become an important
and vibrant field since the pioneering work of Verd{\'u} \cite{verdu}.
The optimal multiuser detector has been proposed by Verd{\'u} in
\cite{verdu86}, however, prohibitive complexity makes its deployment
infeasible and motivated the development of several suboptimal schemes
that are amenable to implementation: The linear \cite{lupas} and
decision feedback \cite{falconer} receivers, the successive
interference canceler~(SIC) \cite{patel}, and the parallel interference
canceler~(PIC) \cite{varanasi}. These receivers require the estimation
of various parameters in order to carry out interference suppression.

In most practical scenarios, the parameter estimation of the
receiver has to be computed adaptively in order to track the
time-varying multipath channel conditions. Several adaptive
receivers have been reported in \cite{rapajic,madhow,miller} and
have proven to be very valuable techniques for interference
mitigation. Specifically for uplink scenarios, SIC
\cite{patel,cho,shynk} and PIC
\cite{varanasi,latva,divsalar,xie,hamouda,li&hamouda} receivers,
which are relatively simple and perform interference
cancellation~(IC) by sequentially or iteratively removing multiple
access interference (MAI), are known to provide significant gains
over RAKE and linear detectors. The works on SIC and PIC detectors
present severe limitations with respect to the amount of
interference to be estimated and cancelled in dispersive and
dynamic environments. This is because SIC and PIC detectors rely
on an amplitude estimate of the parameters to be cancelled which
affect the whole IC procedure.

This work proposes a unified approach to joint adaptive IC
detectors for DS-CDMA systems in frequency selective channels. A
novel framework in which the IC problem is formulated as an
optimization problem of an IC parameter vector for each user and a
unification of existing IC under the same model is described. A
joint optimization method for estimating the IC parameter vector,
the linear receiver front-end filter and the channel parameters is
proposed along with MMSE expressions. Based on the proposed joint
optimization approach, we also present low complexity SG
algorithms for estimating the desired parameters. Simulations for
the uplink of a DS-CDMA system show that the proposed methods
significantly outperform the best known IC receivers.

The rest of this article is organized as follows:
Section~\ref{sec:format} describes a synchronous DS-CDMA system
model. In Section~\ref{sec:pagestyle}, we present the novel
unified framework for IC. Section~\ref{sec:typestyle} details the
proposed joint optimization approach and describes the MMSE
expressions for channel estimation, receiver filter and IC
parameter vector estimation.  Section~\ref{sec:majhead} is devoted
to the proposed SG adaptive algorithms for joint IC and parameter
estimation. Section~\ref{sec:print} is dedicated to the
simulations and to the discussion of the results, whereas the
conclusions are drawn in Section~\ref{sec:page}.

\vspace*{-0.0em} \mysection{DS-CDMA System Model}
\label{sec:format}

Let us consider the uplink of a symbol synchronous quadrature
phase-shift keying~(QPSK) DS-CDMA system with $K$ users, $N$ chips
per symbol and $L_{p}$ propagation paths. Note that a synchronous
model is assumed for simplicity, although it captures most of the
features of asynchronous models with small to moderate delay
spreads. The baseband signal transmitted by the $k$-th active user
to the base station is given by
\begin{equation}
x_{k}(t)=A_{k}\sum_{i=-\infty}^{\infty}b_{k}[i]s_{k}(t-iT),
\end{equation}
where $b_{k}[i] \in \{\pm1 \pm j\}$ with $j^2=-1$ denotes the
$i$-th symbol for user $k$, the real-valued spreading waveform and
the amplitude associated with user $k$ are $s_{k}(t)$ and $A_{k}$,
respectively. The spreading waveforms are expressed by { $s_{k}(t)
= \sum_{i=1}^{N}a_{k}[i]\phi(t-iT_{c})$}, where { $a_{k}[i]\in
\{\pm1/\sqrt{N} \}$}, { $\phi(t)$} is the chip waveform, $T_{c}$
is the chip duration and $N=T/T_{c}$ is the processing gain.
Assuming that the receiver is synchronized with the main path, the
complex envelope of the coherently demodulated composite received
signal is
\begin{equation}
r(t)= \sum_{k=1}^{K}\sum_{l=0}^{L_{p}-1}
h_{k,l}(t)x_{k}(t-\tau_{k,l})+n(t),
\end{equation}
where $h_{k,l}(t)$ and $\tau_{k,l}$ are, respectively, the channel
coefficient and the delay associated with the $l$-th path and the
$k$-th user. Assuming that $\tau_{k,l} = lT_{c}$, the channel
 is constant during each packet transmission, $L_p \leq N$ and the
spreading codes are repeated from symbol to symbol, the received
signal $r(t)$ after filtering by a chip-pulse matched filter and
sampled at chip rate yields the $M$-dimensional received vector
\begin{equation}
\begin{split}
{\bf r}[i] & =  \sum_{k=1}^{K} A_{k} \Big( b_k[i-1]{\bf C}_{k}^p +
{b}_{k}[i] {\bf C}_{k}  \\ & \quad + b_k[i+1]{\bf C}_{k}^s \Big)
{\bf h}_k   + {\bf n}[i],
\end{split}
\end{equation}
where $M=N+L_{p}-1$, ${\bf n}[i] = [n_{1}[i] ~\ldots~n_{M}[i]]^{T}$ is
the complex Gaussian noise vector with $\E[{\bf n}[i]{\bf n}^{H}[i]] =
\sigma^{2}{\bf I}$, where $(\cdot)^{T}$ and $(\cdot)^{H}$ denote
transpose and Hermitian transpose, respectively, $\E[\cdot]$ stands
for expected value, the amplitude of user $k$ is $A_{k}$, ${\bf s}_{k}
= [a_{k}(1) \ldots a_{k}(N)]^{T}$ is the signature sequence for the
$k$-th user, the $M\times L_{p}$ constraint matrices ${\bf C}_{k}^p$,
${\bf C}_{k}$, ${\bf C}_{k}^s$ that contains one-chip shifted versions
of the signature sequence for user $k$ and the $L_{p}\times 1$ vector
${\bf h}_k$ with the discrete-time multipath components are
described by
\begin{equation}
{\bf C}_{k} = \left[\begin{array}{c c c }
a_{k}(1) &  & {\bf 0} \\
\vdots & \ddots & a_{k}(1)  \\
a_{k}(N) &  & \vdots \\
{\bf 0} & \ddots & a_{k}(N)  \\
 \end{array}\right],
\ms
 {\bf h}_k=\left[\begin{array}{c} {h}_{k,0}
\\ \vdots \\ {h}_{k,L_{p}-1}\\  \end{array}\right],
\end{equation}
\begin{equation}
{\bf C}_{k}^p = \left[\begin{array}{c}
{\bf C}_{k,~[N+1:M,1:L_p]} \\
{\bf 0}_{N \times L_p}  \\
 \end{array}\right],
\ms
{\bf C}_{k}^s = \left[\begin{array}{c}
 {\bf 0}_{N \times L_p} \\
{\bf C}_{k,~[1:L_p-1,1:L_p]}  \\
 \end{array}\right],
\end{equation}
where the matrices ${\bf C}_{k}^p$ and ${\bf C}_{k}^s$ account for
the intersymbol interference from the previous and subsequent
symbols, respectively. The subscript $[m:q,j:p]$ denotes the range
of elements of a given matrix used.

\mysection{Unified Framework for Interference Cancellation}
\label{sec:pagestyle}

In this section, we present a unified framework for IC based on
the formulation of the problem as the optimization of an IC
parameter vector for each user. Prior works on IC are heavily
based on the estimation of an amplitude of a user (or an amount of
interference) to be reconstructed and cancelled. This approach is
very sensitive and tends to be inaccurate in multipath scenarios.
Unlike prior works, the proposed approach trades off an amplitude
estimate against a parameter vector estimate for each stage,
incorporating more degrees of freedom to the problem and which
reveals to be significantly more effective than the best known
methods. With this new formulation, we provide a unifying
treatment of IC schemes in what follows.

Let us first consider a conventional IC approach, where the
centralized processor aims to reconstruct the detected data and
subtract them from the $M \times 1$ received data
\begin{equation}
\begin{split}
{\bf r}^{m}_k[i] & = {\bf r}[i] - \sum_{j \in
{\boldsymbol{\mathcal{G}}}}
\hat{A}_{j}^m \Big( \hat{b}_j^m[i-1]{\bf C}_{j}^p + \hat{b}_{j}^m[i] {\bf C}_{j}  \\
& \quad + \hat{b}_j^m[i+1]{\bf C}_{j}^s \Big) \hat{\bf h}_j^m[i]  \\
& = {\bf r}[i] - \sum_{j \in {\boldsymbol{\mathcal{G}}}}
\hat{A}_{j}^m {\bf F}_j^m[i]\hat{\bf h}_j^m[i],
\end{split}
\label{eq:receiverfirst}
\end{equation}
where ${\bf F}_j^m[i] = \Big( \hat{b}_j^m[i-1]{\bf C}_{j}^p +
\hat{b}_{j}^m[i] {\bf C}_{j}  + \hat{b}_j^m[i+1]{\bf C}_{j}^s
\Big)$ is an $M \times L_p$ matrix with the signature code and
symbol estimates of user $j$ at the IC stage $m$,
${\boldsymbol{\mathcal{G}}}= \{ \mathcal{G}_1,~
\mathcal{G}_2,~\ldots,~\mathcal{G}_P \}$ denotes the group of
users with $P$ entries to be reconstructed and subtracted, whereas
the subscript denotes the $k$th user. In the conventional approach
outlined in~\refeq{eq:receiverfirst}, the goal is to compute the
estimates $\hat{A}_{j}^m$, ${\bf F}_j^m[i]$ and $\hat{\bf
h}_k^m[i]$ and this often leads to inaccurate IC for systems in
dispersive channels.

The novel approach we detail here corresponds to a mathematical
reformulation of~\refeq{eq:receiverfirst} and the introduction of an IC parameter
vector
$\boldsymbol{\lambda}_k^m=\big[\lambda_{k,1}^m~\lambda_{k,2}^m~\ldots~
\lambda_{k,P}^m\big]^T$ for each stage $m$ and user $k$, as
expressed by
\begin{equation}
\begin{split}
{\bf r}^{m}_k[i] & = {\bf r}[i] - \sum_{j \in
{\boldsymbol{\mathcal{G}}}}
{\lambda}_{j,k}^m {\bf F}_{j}^m[i]\hat{\bf h}_j^m[i] \\
& = {\bf r}[i] -   {\bf
D}_{\boldsymbol{\mathcal{G}}}^m[i]\boldsymbol{\lambda}_{k}^m[i],
\end{split}
\label{eq:oldeq7}
\end{equation}
where the $M \times P$ matrix with signature codes, symbols and
channels estimates of the group of users
$\boldsymbol{\mathcal{G}}$ is given by
\begin{equation}
{\bf D}_{\boldsymbol{\mathcal{G}}}^m[i] =
\boldsymbol{\mathcal{C}}_{\rm T}^p \boldsymbol{\mathcal{H}}^m
{\bf B}^m[i-1] + \boldsymbol{\mathcal{C}}_{\rm T}
\boldsymbol{\mathcal{H}}^m {\bf B}^m[i] +
\boldsymbol{\mathcal{C}}_{\rm T}^s \boldsymbol{\mathcal{H}}^m {\bf
B}^m[i+1],
\end{equation}
\begin{equation}
\boldsymbol{\mathcal{C}}_{\rm T} = \Big[ {\bf C}_{\mathcal{G}_1} ~
{\bf C}_{\mathcal{G}_2} ~ \ldots ~ {\bf C}_{\mathcal{G}_P} \Big],
~~~ \boldsymbol{\mathcal{C}}_{\rm T}^s = \Big[ {\bf
C}_{\mathcal{G}_1}^s~ {\bf C}_{\mathcal{G}_2}^s ~ \ldots ~ {\bf
C}_{\mathcal{G}_P}^s, \Big],
\end{equation}
\begin{equation}
\boldsymbol{\mathcal{C}}_{\rm T}^p = \Big[ {\bf
C}_{\mathcal{G}_1}^p ~ {\bf C}_{\mathcal{G}_2}^p ~ \ldots {\bf
C}_{\mathcal{G}_P}^p\Big], ~ \boldsymbol{\mathcal{H}}^m = {\rm
diag} \big({\bf h}_{\mathcal{G}_1}^m~{\bf h}_{\mathcal{G}_2}^m ~
\ldots ~ {\bf h}_{\mathcal{G}_P}^m \Big),
\end{equation}
\begin{equation}
{\bf B}^m[i] = {\rm diag} \Big( b_{\mathcal{G}_1}^m[i],~
b_{\mathcal{G}_2}^m[i],~\ldots,~b_{\mathcal{G}_P}^m[i] \Big),
\end{equation}
The matrix ${\bf D}_\mathcal{G}^m[i]$ corresponds to the reconstructed
data of users which belong to group $\boldsymbol{\mathcal{G}}$ and it
is a function of the $M \times (P \cdot L_p)$ code matrices
$\boldsymbol{\mathcal{C}}_{\rm T}^p$, $\boldsymbol{\mathcal{C}}_{\rm
  T}$, $\boldsymbol{\mathcal{C}}_{\rm T}^s$, the $P \times P$ user's
symbol matrix ${\bf B}^m[i]$ and the $(P \cdot L_p) \times P$
channel matrix $\boldsymbol{\mathcal{H}}^m$ with a block diagonal
structure, with all the parameters of the users to be
reconstructed and cancelled.

The mathematical framework detailed in~\refeq{eq:oldeq7} can be
actually used to describe IC schemes which perform SIC and PIC as
particular cases.  For instance, if the designer chooses to detect the
users according to a decreasing power ordering, we obtain the
following SIC approach
\begin{equation}
\label{eq:oldeq12}
\begin{split}
{\bf r}^{m}_k[i] & = {\bf r}[i] - \sum_{j =1}^{k-1}
{\lambda}_{j,k}^m {\bf F}_j^m[i]\hat{\bf h}_j^m[i] \\
& = {\bf r}[i] -   {\bf
D}_{\boldsymbol{\mathcal{G}}_{k-1}}^m[i]\boldsymbol{\lambda}_{k}^m[i],~~~
m =1
\end{split}
\end{equation}
where $\boldsymbol{\mathcal{G}}_{k-1} = \{\mathcal{G}_1, ~
\mathcal{G}_2,~\ldots,~\mathcal{G}_{k-1}\}$ denotes the group of
users to be reconstructed according to the SIC approach
(decreasing power order). Note that there is only one stage
($m=1$) for SIC, user $k$ is detected and the previously detected
users are regenerated and subtracted from the received data ${\bf
r}[i]$ and this is repeated for the remaining users.

Another detection strategy which can be carried out as a
particular case of the framework in~\refeq{eq:oldeq7} is the
multistage detection or PIC. Following the proposed detection
scheme users are detected on the basis of the following structure
\begin{equation}
\begin{split}
{\bf r}^m_k[i] & = {\bf r}[i] - \sum_{\underset{j \neq k}{j=1}}^{K}
{\lambda}_{j,k}^m {\bf F}_j^m[i]\hat{\bf h}_j^m[i] \\
& = {\bf r}[i] -   {\bf
D}^m_{\boldsymbol{\mathcal{G}}_{K-1}}[i]\boldsymbol{\lambda}_{k}^m,~~~
m=1,~2,~\ldots
\end{split}
\end{equation}
where $k$ represents the desired user to be detected, $m$ is the
stage, and ${\boldsymbol{\mathcal{G}}_{K-1}}$ contains all but
the desired $k$th user.

\begin{figure}[t]
  \begin{center}
    \setlength{\unitlength}{0.00078333in}
\begingroup\makeatletter\ifx\SetFigFont\undefined%
\gdef\SetFigFont#1#2#3#4#5{%
  \reset@font\fontsize{#1}{#2pt}%
  \fontfamily{#3}\fontseries{#4}\fontshape{#5}%
  \selectfont}%
\fi\endgroup%
{\renewcommand{\dashlinestretch}{30}
\begin{picture}(3924,2214)(0,-10)
\put(1512,1437){\makebox(0,0)[b]{{\SetFigFont{11}{13.2}{\rmdefault}{\mddefault}{\updefault}$\boldsymbol{w}_{k}[i]$}}}
\path(12,1812)(687,1812)
\blacken\path(567.000,1782.000)(687.000,1812.000)(567.000,1842.000)(567.000,1782.000)
\path(687,387)(312,387)(312,1587)(687,1587)
\blacken\path(567.000,1557.000)(687.000,1587.000)(567.000,1617.000)(567.000,1557.000)
\path(2487,387)(1737,387)
\blacken\path(1857.000,417.000)(1737.000,387.000)(1857.000,357.000)(1857.000,417.000)
\path(2862,1587)(3012,1587)(3012,1887)(3162,1887)
\path(2937,1737)(3087,1737)
\path(3312,1737)(3912,1737)
\blacken\path(3792.000,1707.000)(3912.000,1737.000)(3792.000,1767.000)(3792.000,1707.000)
\path(3687,1737)(3687,387)(3537,387)
\blacken\path(3657.000,417.000)(3537.000,387.000)(3657.000,357.000)(3657.000,417.000)
\path(3687,1137)(2187,1137)(2187,612)(1737,612)
\blacken\path(1857.000,642.000)(1737.000,612.000)(1857.000,582.000)(1857.000,642.000)
\path(2712,1437)(3312,1437)(3312,2037)
	(2712,2037)(2712,1437)
\put(2592,117){\arc{210}{1.5708}{3.1416}}
\put(2592,657){\arc{210}{3.1416}{4.7124}}
\put(3432,657){\arc{210}{4.7124}{6.2832}}
\put(3432,117){\arc{210}{0}{1.5708}}
\path(2487,117)(2487,657)
\path(2592,762)(3432,762)
\path(3537,657)(3537,117)
\path(3432,12)(2592,12)
\put(792,117){\arc{210}{1.5708}{3.1416}}
\put(792,657){\arc{210}{3.1416}{4.7124}}
\put(1632,657){\arc{210}{4.7124}{6.2832}}
\put(1632,117){\arc{210}{0}{1.5708}}
\path(687,117)(687,657)
\path(792,762)(1632,762)
\path(1737,657)(1737,117)
\path(1632,12)(792,12)
\put(792,1392){\arc{210}{1.5708}{3.1416}}
\put(792,2082){\arc{210}{3.1416}{4.7124}}
\put(2157,2082){\arc{210}{4.7124}{6.2832}}
\put(2157,1392){\arc{210}{0}{1.5708}}
\path(687,1392)(687,2082)
\path(792,2187)(2157,2187)
\path(2262,2082)(2262,1392)
\path(2157,1287)(792,1287)
\put(87,1887){\makebox(0,0)[b]{{\SetFigFont{11}{13.2}{\rmdefault}{\mddefault}{\updefault}$\boldsymbol{r}[i]$}}}
\put(1212,312){\makebox(0,0)[b]{{\SetFigFont{11}{13.2}{\rmdefault}{\mddefault}{\updefault}Cancellation}}}
\put(1212,87){\makebox(0,0)[b]{{\SetFigFont{11}{13.2}{\rmdefault}{\mddefault}{\updefault}$\boldsymbol{\lambda}_{k}[i]$}}}
\put(3012,87){\makebox(0,0)[b]{{\SetFigFont{11}{13.2}{\rmdefault}{\mddefault}{\updefault}$\hat{\boldsymbol{h}}_{k}[i]$}}}
\put(87,1062){\makebox(0,0)[b]{{\SetFigFont{11}{13.2}{\rmdefault}{\mddefault}{\updefault}$\boldsymbol{r}_{k}[i]$}}}
\put(3762,1887){\makebox(0,0)[b]{{\SetFigFont{11}{13.2}{\rmdefault}{\mddefault}{\updefault}$\hat{b}_{k}[i]$}}}
\put(2112,87){\makebox(0,0)[b]{{\SetFigFont{11}{13.2}{\rmdefault}{\mddefault}{\updefault}$\hat{\boldsymbol{h}}_{k}[i]$}}}
\put(2487,1887){\makebox(0,0)[b]{{\SetFigFont{11}{13.2}{\rmdefault}{\mddefault}{\updefault}$x_{k}[i]$}}}
\put(1212,537){\makebox(0,0)[b]{{\SetFigFont{11}{13.2}{\rmdefault}{\mddefault}{\updefault}Interference}}}
\put(3012,537){\makebox(0,0)[b]{{\SetFigFont{11}{13.2}{\rmdefault}{\mddefault}{\updefault}Channel}}}
\put(3012,312){\makebox(0,0)[b]{{\SetFigFont{11}{13.2}{\rmdefault}{\mddefault}{\updefault}Estimation}}}
\put(1512,1887){\makebox(0,0)[b]{{\SetFigFont{11}{13.2}{\rmdefault}{\mddefault}{\updefault}Linear Interference}}}
\put(1512,1662){\makebox(0,0)[b]{{\SetFigFont{11}{13.2}{\rmdefault}{\mddefault}{\updefault}Suppression}}}
\path(2262,1737)(2712,1737)
\blacken\path(2592.000,1707.000)(2712.000,1737.000)(2592.000,1767.000)(2592.000,1707.000)
\end{picture}
}
    \vspace*{1em}
    \mycaption{Proposed receiver structure.}
    \label{fig:system}
  \end{center}
\end{figure}

\mysection{Joint Interference Cancellation and Parameter Estimation
  Method }
\label{sec:typestyle}

In this section, we present a novel strategy for joint interference
cancellation and parameters estimation based on the formulation given
in the previous section. The idea is to consider the problems of
receiver filter, channel, and IC parameter vector estimation jointly
and devise MMSE expressions to solve it.  Specifically, we consider
here that a linear receiver filter is employed at the front-end of the
IC detector. The proposed receiver structure is showed in
Figure~\ref{fig:system}.  Let us first consider the following cost
functions
\begin{equation}
\label{eq:old14}
J_1({{\bf w}_k^m[i]}) = \E \Big[|b_k[i] - {\bf
w}_k^{m,H}[i]{\bf r}_k^m[i]|^2 \Big],
\end{equation}
\begin{equation}
\label{eq:old15}
\begin{split}
J_2({\boldsymbol{\lambda}_k^m[i], ~ \hat{\bf h}_k^m[i]}) & =  \E
\Big[||{\bf F}_k^m[i] \hat{\bf h}_{k}^m[i] - {\bf r}[i] + {\bf
D}^m_{\boldsymbol{\mathcal{G}}}[i]\boldsymbol{\lambda}_{k}^m[i]
||^2 \Big].
\end{split}
\end{equation}
By minimizing~\refeq{eq:old14} with respect to the receiver linear filter
${\bf w}_k^m[i]$ we obtain the Wiener-Hopf-like expressions
\begin{equation}
\label{eq:old16}
{\bf w}_k^m[i] = {\bf R}^{-1}_{{\bf r}_k^m}[i]{\bf p}_{b_k}[i],
\end{equation}
where ${\bf R}_{{\bf r}_k^m}[i] = \E\big[{\bf r}_k^m[i]{\bf
  r}_k^{m,H}[i]\big]$ is the $M\times M$ covariance matrix and ${\bf
  p}_{b_k}[i] = \E\big[b_k^*[i]{\bf r}_k^m[i]\big]$ is the $M \times 1$
cross-correlation vector. By taking the gradient terms
of~\refeq{eq:old15} with respect to the IC parameter vector
$\boldsymbol{\lambda}_k^m[i]$ and equating them to zero we get
\begin{equation}
\label{eq:old17} \boldsymbol{\lambda}_k^m[i] = {\bf R}^{-1}_{{\bf
D}^m_{\boldsymbol{\mathcal{G}}}}[i]{\bf p}_{{\bf F}_k^m}[i],
\end{equation}
where  ${\bf R}_{{\bf
    D}^m_{\boldsymbol{\mathcal{G}}}}[i] = \E\big[ {\bf
  D}^{m,H}_{\boldsymbol{\mathcal{G}}}[i] {\bf
  D}^m_{\boldsymbol{\mathcal{G}}}[i]  \big]$ is the $P\times P$
  covariance matrix and ${\bf p}_{{\bf F}_k^m}[i] = \E\big[ {\bf
  D}^{m,H}_{\boldsymbol{\mathcal{G}}}[i] ( {\bf r}[i] - {\bf
  F}_k^m[i] \hat{\bf h}_{k}^m[i])\big]$ is the $P \times 1$
cross-correlation vector. By minimizing~\refeq{eq:old15} with
regard to the channel estimate $\hat{\bf h}_k^m[i]$ we obtain the
last system of linear equations
\begin{equation}
\label{eq:old18}
\hat{\bf h}_k^m[i] = {\bf R}^{-1}_{{\bf F}_k^m}[i]{\bf p}_{{\bf
D}^m_{\boldsymbol{\mathcal{G}}}}[i],
\end{equation}
where ${\bf R}_{{\bf F}_k^m}[i] = \E\big[ {\bf
  F}_k^{m,H}[i]{\bf F}_k^m[i]  \big]$
is the $L_p \times L_p$ covariance matrix and ${\bf p}_{{\bf
D}^m_{\boldsymbol{\mathcal{G}}}}[i] = \E\big[ {\bf
  F}_k^{m,H}[i] ({\bf r}[i] - {\bf
  D}^m_{\boldsymbol{\mathcal{G}}}[i]\boldsymbol{\lambda}_{k}^m[i] )
\big]$ is the $P \times 1$ cross-correlation vector. The
expressions in \refeq{eq:old16}-\refeq{eq:old18} are not
closed-form ones as the interference cancellation and the receiver
linear filter parameters depend on the channel and vice-versa.
This means that \refeq{eq:old16}-\refeq{eq:old18} have to be
iterated in order to seek a solution to the optimization problem.
The detected symbols are obtained as follows:
\begin{equation}
\label{eq:detector} {\hat{b}}_{k}[i] = {\rm sgn} \Big[ \Re
\big(x_k[i])\Big] + j \Big[ \Im \big( x_k[i] \big) \Big],
\end{equation}
where $x_k[i]={\bf w}_k^{m,H}[i]{\bf r}_k^m[i]$ and ${\rm
sgn}\big[ \cdot \big]$ is the signum function. The expressions in
\refeq{eq:old16}-\refeq{eq:old18} require a computational
complexity of $O(M^3)$, $O(P^3)$ and $O(L_p^3)$ for the estimation
of ${\bf w}_k^m[i]$, $\boldsymbol{\lambda}_k^m[i]$ and $\hat{\bf
h}_k^m[i]$, respectively.  This cubic cost is due to the required
matrix inversions. Note that the optimization of
$J_2({\boldsymbol{\lambda}_k^m[i], ~ \hat{\bf h}_k^m[i]})$ in
\refeq{eq:old15} seeks to eliminate MAI, whereas the optimization
of $J_1({{\bf w}_k^m[i]})$ in ~\refeq{eq:old15} aims to suppress
the residual MAI and intersymbol interference. However, a problem
is that the statistics are not known a priori and the designer has
to estimate them. In what follows, we seek adaptive SG solutions
based on the joint optimization of the parameters formulated
in~\refeq{eq:old14} and \refeq{eq:old15} for estimating these
parameters and allowing amenable implementation. \vspace*{-0.75em}

\mysection{Adaptive Estimation Algorithms}
\label{sec:majhead}

In this section, we propose adaptive estimation algorithms based on
the joint optimization problems stated in~\refeq{eq:old14} and
\refeq{eq:old15}. In order to develop SG adaptive algorithms, we
compute the instantaneous gradients of~\refeq{eq:old14} with respect
to ${\bf w}_k^m[i]$, and of \refeq{eq:old15} with respect to
$\boldsymbol{\lambda}_k^m[i]$ and $\hat{\bf h}_k^m[i]$ as follows
\begin{equation}
\label{eq:oldeq19}
\begin{split}
\nabla {J_{1_{{\bf w}_k^m[i]}}} = \frac{\partial J_{{\bf
w}_k^m[i]}}{
\partial {\bf w}_k^{*~m}[i]} & = - {\bf r}_k^m[i] \Big( b_k[i] - {\bf w}_k^{m,H}[i]{\bf
r}_k^m[i] \Big)^{*} \\ & = - {\bf r}_k^m[i] e_k^{*~m}[i],
\end{split}
\end{equation}
\begin{equation}
\label{eq:oldeq20}
\begin{split}
\nabla J_{2_{\boldsymbol{\lambda}_k^m[i]}} & = \frac{\partial
J_{\boldsymbol{\lambda}_k^m[i], ~ \hat{\bf h}_k^m[i]}}{
\partial \boldsymbol{\lambda}_k^{*~m}[i]} \\ & =  {\bf D}^{m,H}_{\boldsymbol{\mathcal{G}}}[i] ({\bf
F}_k^m[i] \hat{\bf h}_{k}^m[i] - {\bf r}[i] + {\bf
D}^m_{\boldsymbol{\mathcal{G}}}[i]\boldsymbol{\lambda}_{k}^m[i])
\\
& =  {\bf D}^{m,H}_{\boldsymbol{\mathcal{G}}}[i] ({\bf F}_k^m[i]
\hat{\bf h}_{k}^m[i] - {\bf r}_k^m[i] ) \\
& =  {\bf D}^{m,H}_{\boldsymbol{\mathcal{G}}}[i] {\bf e}_k^{m}[i],
\end{split}
\end{equation}
\begin{equation}
\label{eq:oldeq21}
\begin{split}
\nabla J_{2_{ \hat{\bf h}_k^m[i]}} & = \frac{\partial
J_{\boldsymbol{\lambda}_k^m[i], ~ \hat{\bf h}_k^m[i]}}{
\partial \hat{\bf
h}_k^{m}[i]} \\ & =  {\bf F}_k^{m,H}[i] ({\bf F}_k^m[i] \hat{\bf
h}_{k}^m[i] - {\bf r}[i] +  {\bf
D}^m_{\boldsymbol{\mathcal{G}}}[i]\boldsymbol{\lambda}_{k}^m[i] )
\\ & =  {\bf F}_k^{m,H}[i] ({\bf F}_k^m[i] \hat{\bf
h}_{k}^m[i] - {\bf r}_k^m[i]) \\
& =  {\bf F}_k^{m,H}[i] {\bf e}_k^{m}[i]
\end{split}
\end{equation}
By using the well-known gradient-descent rules and introducing
step sizes, we devise the following SG adaptive estimators:
\begin{equation}
\label{eq:old22}
{\bf w}_k^m[i+1] = {\bf w}_k^m[i] + \mu_w e_k^{*~m}[i] {\bf
r}_k^m[i],
\end{equation}
\begin{equation}
\label{eq:old23} \boldsymbol{\lambda}_k^m[i+1] =
\boldsymbol{\lambda}_k^m[i]- \mu_\lambda {\bf
D}^{m,H}_{\boldsymbol{\mathcal{G}}}[i] {\bf e}_k^{m}[i],
\end{equation}
\begin{equation}
\label{eq:old24} \hat{\bf h}_k^m[i+1] = \hat{\bf h}_k^m[i] -
\mu_h{\bf F}_k^{m,H}[i] {\bf e}_k^{m}[i],
\end{equation}
where $\mu_w$, $\mu_\lambda$, and $\mu_h$ are the step sizes for
the adaptive SG estimators for the receiver filter, the IC
parameter vector, and the channel, respectively, and scalar and
vector errors are given by
\begin{equation}
\label{eq:old25}
e_k^m[i] = b_k[i] - {\bf w}_k^{m,H}[i]{\bf r}_k^m[i],
\end{equation}
\begin{equation}
\label{eq:old26} {\bf e}_k^{m}[i] = {\bf F}_k^m[i] \hat{\bf
h}_{k}^m[i] - {\bf r}[i] +  {\bf
D}^m_{\boldsymbol{\mathcal{G}}}[i]\boldsymbol{\lambda}_{k}^m[i]
\end{equation}
The adaptive algorithms described above should be iterated in
order to converge to a solution since the parameters estimated by
them are inter-dependent.  The complexity of the algorithms in
\refeq{eq:old22}-\refeq{eq:old26} is $O(M)$, $O(PM)$ and $O(ML_p)$
for the estimation of the receiver filter ${\bf w}_k^m[i]$, the IC
parameter vector $\boldsymbol{\lambda}_k^m[i]$, and the channel
$\hat{\bf h}_k^m[i]$, respectively.

\mysection{Simulations} \label{sec:print}

In this section, we evaluate the bit error rate (BER) performance
of the proposed joint interference cancellation and parameter
estimation algorithms. We compare the proposed algorithms with the
best known methods of interference mitigation, namely, the linear,
the SIC, and the PIC detectors using SG algorithms as the
estimation procedure. The DS-CDMA system employs randomly
generated spreading sequences of length $N=16$. The channels ${\bf
h}_k = [h_{k,0} ~ h_{k,1} ~\ldots ~h_{k,{L_{p}-1}}]^T$ are modeled
by a tapped-delay-line with $L_p=9$ taps and are normalized such
that $\sum_{l=0}^{L_{p}-1}|h_l|^2 = 1$. Specifically, of the
$L_{p}=9$ taps, there are only $3$ non-zero paths with complex
random gains, whose real and imaginary parts are generated in each
transmitted packet by uniform continuous random variables~(r.v.)
between $-1$ and $1$. The spacing between the paths in each packet
is given by a discrete uniform r.v.\ between $1$ and $3$ chips.
The system has a power distribution among the users for each trial
that follows a log-normal distribution with associated standard
deviation of $3$ dB, the performance is shown in terms of average
$E_b/N_0$, and all curves are averaged over $100$ runs. The
packets used in the simulations have $1500$ symbols and the
training sequences have $150$ symbols.  After the training
sequences, the receivers are switched to decision-directed mode.
The receiver filters used in the simulations have $M=N+L_p-1=24$
taps. The step sizes of all algorithms are optimized for each
situation to ensure the best BER performance for each packet. All
the receivers considered in this part employ a linear receiver as
the front-end. When there is no IC after the linear front-end, the
structure corresponds to a conventional adaptive linear receiver
\cite{rapajic}-\cite{miller}. The proposed jointly optimized~(JO)
algorithms with SIC and PIC receivers are denoted JO-SIC and
JO-PIC, respectively. For the PIC receivers we used $m=3$ stages
and the amplitude estimation based on the estimate at the output
of linear receiver front-end \cite{li&hamouda} and for the
proposed JO-PIC we also use $m=3$ stages. For the existing SIC
receivers \cite{cho,shynk} we used the amplitude estimation
algorithm reported in \cite{shynk}.

\begin{figure}[t]
\begin{center}
\def\epsfsize#1#2{1\columnwidth}
\epsfbox{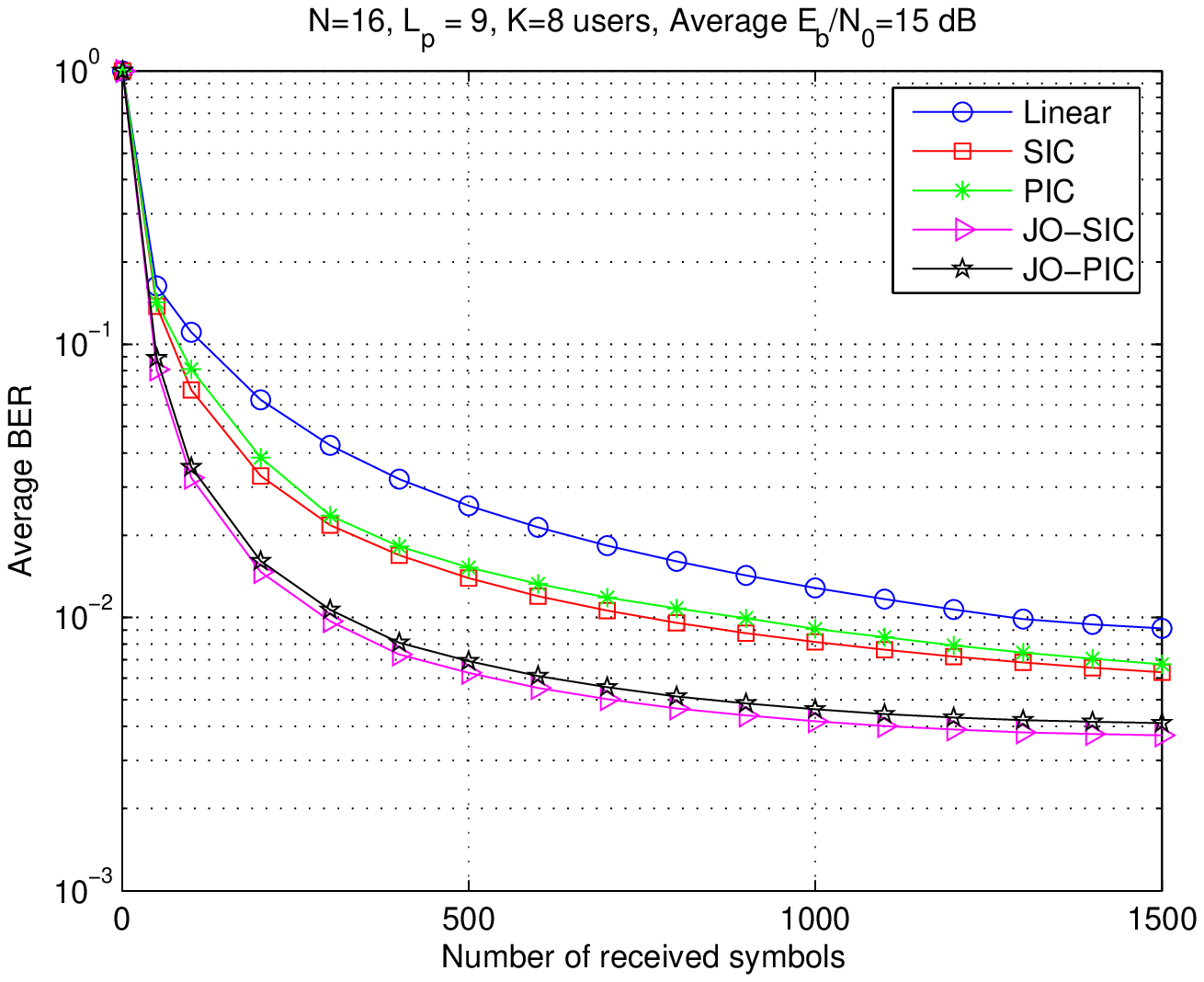} \mycaption{\small BER performance versus number
of received symbols.} \label{fig:berXsymbols}
\end{center}
\end{figure}

In the first experiment, shown in Figure~\ref{fig:berXsymbols}, we
assess the BER convergence performance of the proposed adaptive
estimation algorithms and receiver structures. The results indicate
that the proposed JO-SIC and JO-PIC receivers have the best
performance among the compared structures. The JO-SIC slightly
outperforms the JO-PIC, which is followed by the SIC, the PIC, and the
linear receivers. In particular, we notice that the convergence
performance of the JO-SIC and JO-PIC is significantly superior to the
remaining approaches. This is because the IC is substantially more
accurate than the IC carried out by the existing SIC and PIC
approaches.

\begin{figure}[t]
\begin{center}
\def\epsfsize#1#2{1\columnwidth}
\epsfbox{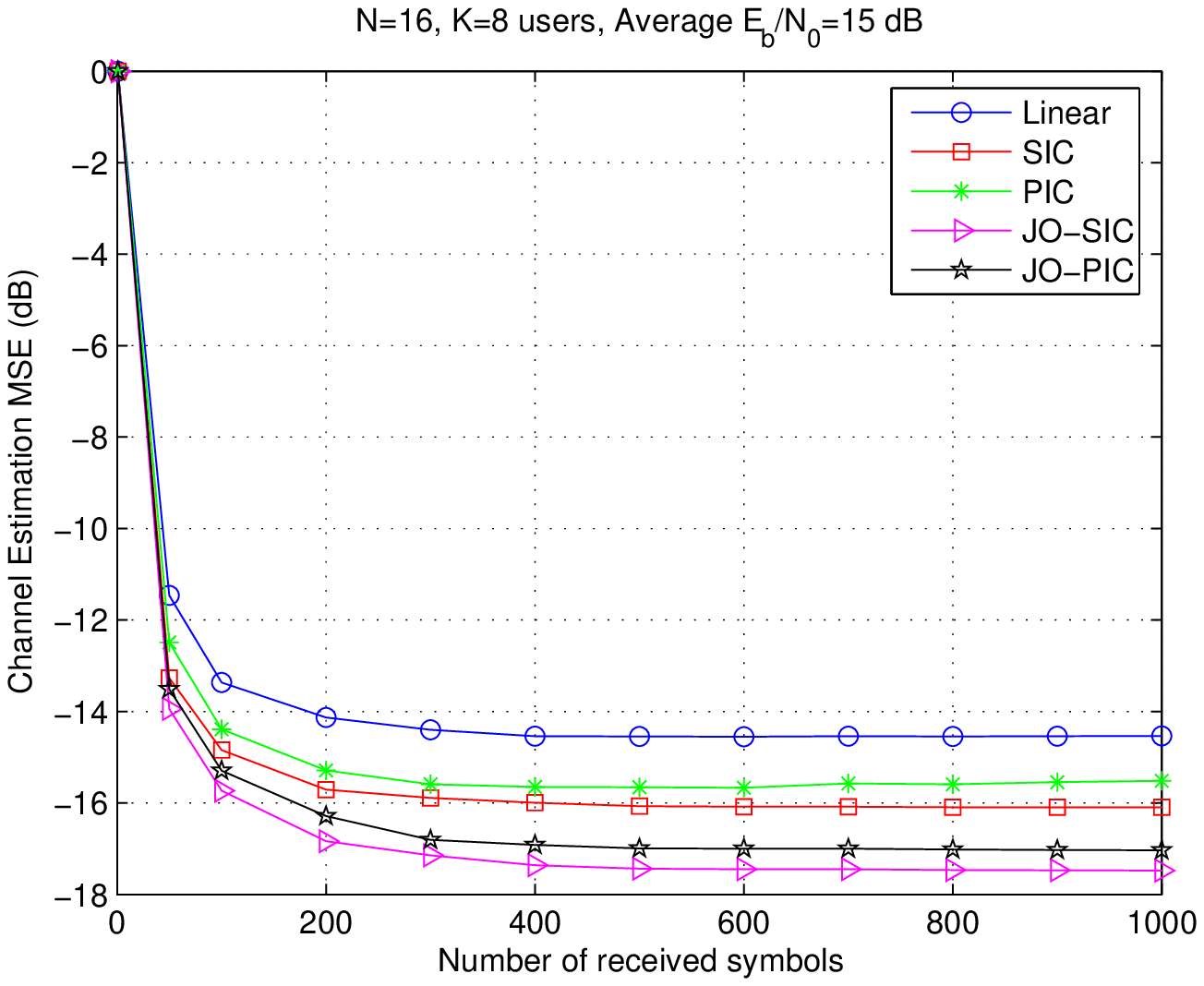} \mycaption{\small MSE performance of channel
estimator versus number of received symbols.}
\label{fig:mseXsymbols_channel}
\end{center}
\end{figure}

The second experiment, depicted in
Figure~\ref{fig:mseXsymbols_channel}, evaluates the MSE performance of
the channel estimators. The channel estimators of IC schemes clearly
benefit by the cancellation process as compared to the linear
estimator without IC. The use of IC for improving the performance is
particularly relevant for the proposed JO-SIC and JO-PIC, which
achieve the best performance.

\begin{figure}[t]
\begin{center}
\def\epsfsize#1#2{1\columnwidth}
\epsfbox{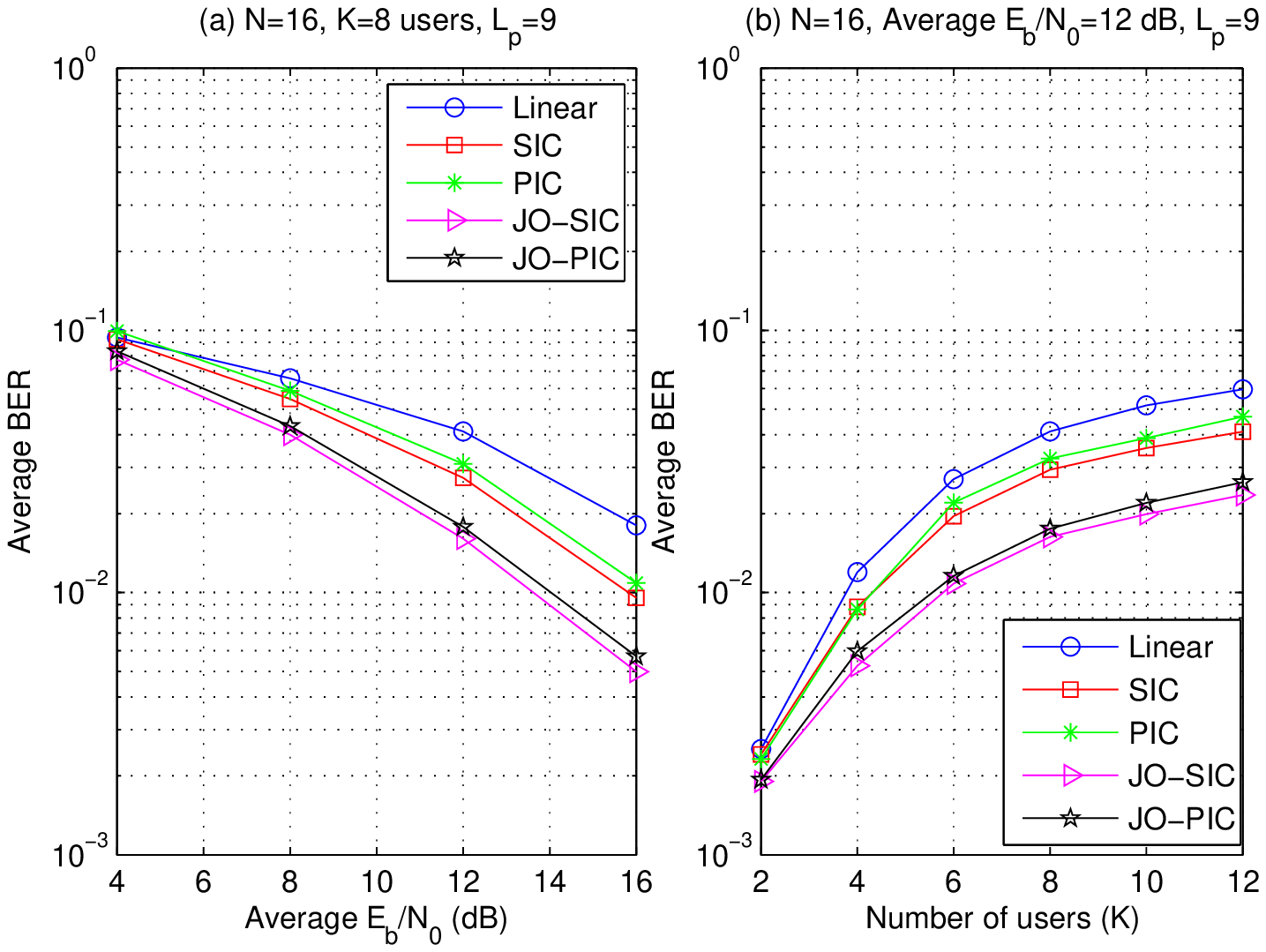} \mycaption{\small BER performance of receivers
versus (a) $E_b/N_0$ and (b)~number of users (K).}
\label{fig:berXser&users}
\end{center}
\end{figure}

The BER versus $E_b/N_0$ and number of users is illustrated in
Figure~\ref{fig:berXser&users}. The curves indicate that the best
performance is obtained by the proposed JO-SIC and JO-PIC receiver and
algorithms. The plots show that the proposed JO-SIC and JO-PIC
receivers can save up to $4$ dB in $E_b/N_0$ for the same BER as
compared with the linear receiver, and up to $2.5$ dB as compared with
existing SIC and PIC detectors. In terms of system capacity, the
proposed detectors and algorithms provide a substantial capacity
improvement and can accommodate up to $50\%$ more users for this small
system.

\mysection{Conclusions}
\label{sec:page}

This work proposed a unified framework for IC in DS-CDMA systems,
which formulates the IC problem as the optimization of an IC
parameter vector. A joint optimization method for estimating the
IC parameter vector, the receiver linear front-end filter and the
channel parameters along with MMSE expressions was also presented.
Low-complexity adaptive estimation algorithms were developed for
jointly estimating the desired parameters. The results for the
uplink of a synchronous DS-CDMA system show that the proposed
methods significantly outperform the best known IC receivers.

\bibliographystyle{IEEEbib}
\bibliography{strings,refs}

\vspace*{-1em}

\begin{thebibliography}{100}

{
\small

\bibitem{verdu}
S. Verdu, {\it Multiuser Detection}, Cambridge, 1998.
\vspace*{-0.5em}

\bibitem{verdu86}
S. Verdu, ``Minimum Probability of Error for Asynchronous Gaussian
Multiple-Access Channels", {\it IEEE Trans. Info. Theory}, vol.
IT-32, no. 1, pp. 85-96, Jan., 1986. \vspace*{-0.5em}

\bibitem{lupas}
R. Lupas and S. Verdu, ``Linear multiuser detectors for
synchronous code-division multiple-access channels," \textit{ IEEE
Trans. Inform. Theory}, vol. 35, pp. 123–136, Jan., 1989.
\vspace*{-0.5em}


\bibitem{falconer}
M. Abdulrahman, A. U. K. Sheikh, and D. D. Falconer, ``Decision
Feedback Equalization for CDMA in Indoor Wireless Communications,"
{\it IEEE Journal on Selected Areas in Communications}, vol 12,
no. 4, May, 1994. \vspace*{-0.5em}


\bibitem{patel}
P. Patel and J. Holtzman, ``Analysis of a Simple Successive
Interference Cancellation Scheme in a DS/CDMA Systems", \textit{
IEEE Journal on Selected Areas in Communications}, vol. 12, no. 5,
June, 1994. \vspace*{-0.5em}


\bibitem{varanasi}
M. K. Varanasi and B. Aazhang, ``Multistage detection in
asynchronous CDMA communications," \textit{ IEEE Transactions on
Communications}, vol. 38, no. 4, pp. 509-19, April, 1990.
\vspace*{-0.5em}


\bibitem{rapajic}
P. B Rapajic and B. S. Vucetic, ``Adaptive receiver structures for
asynchronous CDMA systems", \textit{IEEE JSAC}, vol. 12, no. 4,
pp. 685-697, May 1994. \vspace*{-0.5em}


\bibitem{madhow} U. Madhow and M. L. Honig, ``MMSE
interference suppression for direct-sequence spread-spectrum
CDMA," \textit{ IEEE Transactions on Communications}, vol. 42, no.
12, pp. 3178-88, December, 1994. \vspace*{-0.5em}


\bibitem{miller} M. L. Honig, S.
L. Miller, M. J. Shensa, and L. B. Milstein, ``Performance of
Adaptive Linear Interference Suppression in the Presence of
Dynamic Fading ," {\it IEEE Trans. on Communications}, vol. 49,
no. 4, April 2001. \vspace*{-0.5em}


\bibitem{cho}
Y. Cho and J. H. Lee,``Analysis of an adaptive SIC for near–far
resistant DS-CDMA," \textit{IEEE Trans. on Commun.}, vol. 46, no.
11, November 1998. \vspace*{-0.5em}


\bibitem{shynk}
K. C. Lai and J. J. Shynk, ``Steady-State Analysis of the Adaptive
Successive Interference Canceler for DS/CDMA Signals",
\textit{IEEE Trans. on Sig. Proc.}, vol. 49, no. 10, October,
2001. \vspace*{-0.5em}


\bibitem{latva}
M. Latva-aho and J. Lilleberg, ``Parallel Interference
Cancellation in Multiuser CDMA Channel Estimation,"
\textit{Wireless Personal Communications}, vol. 7, pp. 171-195,
August 1998. \vspace*{-1em}


\bibitem{divsalar}
D. Divsalar, M. K. Simon, and D. Raphaeli, ``Improved parallel
interference cancellation for CDMA," \textit{IEEE Trans. on
Communications}, vol. 46, no. 2, pp. 258-68, February 1998.
\vspace*{-0.5em}


\bibitem{xie}
G. Xie, J. Weng, T. Le-Ngoc, and S. Tahar, ``Adaptive multistage
parallel interference cancellation for CDMA," \textit{ IEEE JSAC},
vol. 17, no. 10, pp. 1815-27, October 1999. \vspace*{-0.5em}


\bibitem{hamouda}
W. A. Hamouda and P. J. McLane, ``A Fast Adaptive Algorithm for
MMSE Receivers in DS-CDMA Systems"," \textit{IEEE Sig. Proc.
Letters}, vol. 11, no. 4, February 2004. \vspace*{-0.5em}

\bibitem{li&hamouda}
M. Li and W.Hamouda, ``Adaptive Multistage Detection for DS-CDMA
Systems in Multipath Fading Channels,"  \textit{Proc. IEEE
VTC-Spring 2005}, May 2005, Sweden. \vspace*{-0.5em}


\bibitem{delamare_mber}
R. C. de Lamare, R. Sampaio-Neto, ``Adaptive MBER decision feedback
multiuser receivers in frequency selective fading channels",
\textit{ IEEE Communications Letters}, vol. 7, no. 2, Feb. 2003, pp.
73 - 75.\vspace*{-0.5em}

\bibitem{delamare_spadf}
R. C. de Lamare, R. Sampaio-Neto, ``Minimum Mean-Squared Error
Iterative Successive Parallel Arbitrated Decision Feedback Detectors
for DS-CDMA Systems", IEEE Trans. on Communications, vol. 56, no. 5,
May 2008, pp. 778 - 789. \vspace*{-0.5em}

\bibitem{reuter}
M. Reuter, J.C. Allen, J. R. Zeidler, R. C. North, ``Mitigating
error propagation effects in a decision feedback equalizer",
\textit{IEEE Trans. on Communications}, vol. 49, no. 11, November
2001, pp. 2028 - 2041.

\bibitem{delamare_itic}
R. C. de Lamare, R. Sampaio-Neto, A. Hjorungnes, ``Joint iterative
interference cancellation and parameter estimation for cdma
systems", \textit{IEEE Comm. Letters}, vol. 11, no. 12, Dec. 2007,
pp. 916 - 918.\vspace*{-0.5em}

\bibitem{stspadf}
Y. Cai and R. C. de Lamare, "Adaptive Space-Time Decision Feedback
Detectors with Multiple Feedback Cancellation", \textit{IEEE
Transactions on Vehicular Technology}, vol. 58, no. 8,  October
2009, pp. 4129 - 4140. \vspace*{-0.5em}

\bibitem{FL11}
R. Fa, R. C. de Lamare, ``Multi-Branch Successive Interference
Cancellation for MIMO Spatial Multiplexing Systems", \textit{ IET
Communications}, vol. 5, no. 4, pp. 484 - 494, March 2011.
\vspace*{-0.5em}

\bibitem{mfsic}
P. Li, R. C. de Lamare, R Fa, "Multiple feedback successive
interference cancellation detection for multiuser MIMO systems",
\emph{IEEE Transactions on  Wireless Communications}, vol. 10, no.
8, 2434-2439, 2011. \vspace*{-0.5em}


\bibitem{mdfpic}
P. Li and R. C. de Lamare, "Adaptive Decision-Feedback Detection
With Constellation Constraints for MIMO Systems", \emph{IEEE
Transactions on Vehicular Technology}, vol. 61, no. 2, 853-859,
2012.\vspace*{-0.5em}
 }

\end{thebibliography}

\end{document}